\documentclass[cameraready]{Interspeech}

\title{Microphone Geometry Aware Regional Target Speaker Extraction}

\author[]{Shitong}{Xu}
\author[]{Niki}{Trigoni}
\author[]{Andrew}{Markham}
\address{Department of Computer Science, University of Oxford, United Kingdom}
\email{\{shitong.xu, niki.trigoni, andrew.markham\}@cs.ox.ac.uk}


\usepackage{comment}
\usepackage{amsmath} 
\usepackage{amssymb} 
\usepackage{bm}  
\usepackage[table]{xcolor}
\usepackage{tabularx}
\usepackage{multirow}
\usepackage{multicol}

\usepackage{listings}
\usepackage{xcolor}
\usepackage{subfigure}

\usepackage{tikz}
\usepackage{pgfplots}

\lstdefinelanguage{json}{
    basicstyle=\ttfamily\scriptsize,
    numbers=left,
    numberstyle=\tiny,
    stepnumber=1,
    numbersep=5pt,
    showstringspaces=false,
    breaklines=true,
    frame=single,
    backgroundcolor=\color{gray!5},
    stringstyle=\color{blue},
    keywordstyle=\color{red},
    commentstyle=\color{green!50!black}
}

\begin{document}
\maketitle

\begin{abstract}
Regional target speaker extraction (RSE) aims to extract speech originating from a specified angular region, relaxing the requirement for precise target speaker location when performing location based speech extraction. 
However, existing RSE methods typically assume a fixed microphone array configuration and microphone number, limiting their applicability to unseen microphone geometries and varying numbers of microphones.
In this work, we propose a geometry-aware RSE framework capable of extract with varying microphone array geometry and microphone numbers. Specifically, we introduce an attention-based multichannel front-end to encode inter-microphone phase information, together with a spatial audio reconstruction-based auxiliary objective to improve the front-end spatial representations quality. Experimental results demonstrate that the proposed framework improves SI-SDRi by 5.1 dB over geometry-aware baselines across microphone arrays with four-microphones, while maintaining a performance gap of less than 0.6 dB compared with a tetrahedral geometry-specific baseline when evaluated on the tetrahedral microphone array.
\end{abstract}

\section{Introduction} \label{sec:introduction}

Target speaker extraction (TSE) aims to isolate the speech of a desired speaker from an audio mixture containing multiple speakers. By leveraging spatial information, prior studies have explored direction-conditioned TSE, where the target speaker is specified by their direction of arrival (DoA). However, these approaches typically require accurate directional information, which may not be readily available in practical scenarios where only an approximate target direction is known.

To address this limitation, recent studies have explored regional target speaker extraction (RSE), which aims to extract speech from a specified angular region while suppressing interfering speakers outside the region. By relaxing the requirement for precise speaker localization, RSE enables more flexible spatially conditioned speech extraction. For instance, an RSE system can capture speech from individuals passing through a wide entrance without requiring their exact locations. 
Existing RSE approaches have explored different encoding strategies of the target angular region. DSENet\cite{dsenet} encodes the target region using its central azimuth direction and angular width. DPARNet-RSE \cite{dparn-rse} represents the region through its left and right azimuth angular boundaries. Other approaches, such as SoundTRC \cite{soundtrc} and FoVNet \cite{fovnet}, discretize the angular space into predefined bins for beamforming, and identify those corresponding to the target region. Despite these advances, existing methods generally assume a fixed number of input microphones, and their robustness to variations in microphone array geometry remains insufficiently explored. Such constraints limit their applicability to previously unseen microphone configurations and under microphone failure. Furthermore, the influence of microphone array geometry on the precision of regional extraction, particularly the ability to distinguish speakers near the target region boundaries, has received limited attention.

In this work, we propose a microphone geometry aware RSE framework that supports varying numbers of microphones and array geometries. Specifically, we introduce an attention-based multichannel front-end that encodes inter-microphone phase information. To encourage the front-end to preserve spatial cues essential for angular discrimination, we introduce a reconstruction-based auxiliary objective that explicitly supervises the learned multichannel representations. We conduct extensive experiments across diverse microphone array geometries and evaluate the proposed framework using real-world audio recordings. Beyond conventional speech extraction metrics, we investigate the spatial selectivity of RSE systems by examining their ability to accurately retain or suppress speakers located near the boundaries of the target angular region. Experimental results demonstrate that the proposed framework outperforms geometry-aware baselines across varying microphone configurations, while achieving competitive performance compared to geometry-specific models on real world audio recordings.

Experiments demonstrate that the proposed framework improves SI-SDRi by 5.1 dB over geometry-aware baselines across four microphone configurations, while maintaining a performance gap of less than 0.6 dB in SI-SDRi compared with the tetrahedral geometry-specific baseline when evaluated on the tetrahedral microphone array.






\section{Related Work} \label{sec:related_work}

\subsection{Location based Target Speaker Extraction}

Location information of the target speaker provide complementary information for the model to extract the speaker's voice in addition to using the audio enrollments \cite{soundcompass, soundtrc}. However, exact target speaker's direction and distance may be hard to obtain in reality. Prior works have explored relaxing the strong spatial requirement to extract the speakers within given angular and distance range. 
DSENet \cite{dsenet} encode the audio received by a tetrahedral microphone array with a 4 channel 1D convolution layer. The target angular region center and width are encoded as masks boardcasted along the time dimension, and fuse with the audio embedding via elementwise-multiplication.
DPARNet-RSE \cite{dparn-rse} encode the microphone position as phase delay based on the target angular region center, and encode the left and right extraction angular boundaries as concatenated sine and cosine values, pass through linear layer to the audio feature dimension, and fuse with the audio embedding by cross attention.

Both DSENet and DPARN-RSE models are optimized in a two-stages manner, where the first stage include only samples where a single speaker exist in the target angular region. The second stage involves more diverse setting where zero to two target speakers exist, training model to extract the overlapped speech of multiple speakers, or output silent audio if no speaker is in the target region. 
ReZero \cite{rezero} and SoundTRC \cite{soundtrc} both optimized the model to perform extraction conditioned on the distance of the speaker from the microphone array. ReZero \cite{rezero} encode the distance information with learnable modules before fuse with the audio embedding via element-wise addition. SoundTRC \cite{soundtrc} in contrast define the target speaker region as vectors. Each element in the target region vector represents the distance threshould of source up to which user want to extract. This representation enables flexible extraction regions with distance boundary varies with direction.

\subsection{Spatial Audio Processing with Arbitrary Microphone Geometries}

In sound source localization field, multiple prior works have explored localization under arbitrary known microphone geometries \cite{gidoaenet, phasecoder}. In comparison to these localization frameworks estimating the source category and direction, our problem formulation requires reconstructing the speech signal satisfying a spatial query.

In the audio separation and extraction field, multple prior works \cite{spatialnet, tac, microsoft-sep, vararray} have leveraged the phase difference from different microphone recordings to assist separation. However, these methods perform inference in a microphone geometry-agnostic manner and do not provide an explicit mechanism for relating a user-specified angular region to the microphone positions.
Although Galatea \cite{Galatea} performed microphone geometry-aware separation and localization, their exploration focused on the scenario where only 1-2 speakers exist in the scene, and lacked evaluation on real-world recordings.

Among the microphone geometry-aware models that perform extraction based on target speaker location,
GC-SSF \cite{gc-ssf} extract conditioned on the exact target speaker's direction. This formulation assumes that the target direction is known and does not directly address extraction from a coarsely specified angular region. In addition, these methods do not allow extracting multiple speakers in an angular region, e.g. to extract the speakers performing on the stage 45 degrees to the front of the receiver. 
In addition, FoVNet \cite{fovnet} resolve the varying microphone number and geometry structure by beamforming into pre-defined direction bins, forming fixed number of audio channels for the subsequent speaker extraction modules. In our experiments, approaches based on this discretized beamforming representation exhibit lower extraction performance and weaker simulation-to-real generalization.

In conclusion, prior work has addressed flexible spatial queries and varying microphone geometries largely as separate problems. Region-based extraction methods provide flexible spatial conditioning but typically assume a predefined array configuration, whereas geometry-generalizable localization and separation methods either do not reconstruct speech satisfying a spatial query or do not explicitly associate the query with the array geometry. Our work focuses on the intersection, performing speech extraction from a user-specified angular region under varying microphone numbers and geometries.

\section{Methodology}
\label{sec:methodology}

\subsection{Problem Formulation}

We focus on geometry-aware regional target speech extraction task, where the input is the spatial audio received by a microphone array of arbitrary number of microphone and compact geometry, the coordinate of each microphone, and the azimuth angle range defining the target speaker range. The objective is to extract all the speakers' overlapped reverberant speech originating within that region. We do not set boundary on the target speakers' elevation angle range, and augment our model to extract all the speakers with azimuth angle inside the specified angle range. For simplicity, if without otherwise explained, we refer to the angle range as azimuth angle in the following explanation. 

\subsection{Geometry-Aware Microphone-Count-Agnostic Encoding Frontend}

To extract spatially informative features from microphone arrays with varying geometries and microphone counts, we propose a geometry-aware encoding frontend. The frontend consists of four stages: Per-Channel Feature Encoding, Microphone Geometry Encoding, Cross-Microphone Attention Module, and Cross-Channel Aggregation.

\textbf{Per-Channel Feature Encoding}
Given the multichannel audio signals, we construct a three-channel spectral representation by stacking the real, imaginary, and log-magnitude components of each microphone's STFT, yielding a multichannel audio representation $\mathbf{X}\in\mathbb{R}^{M\times 3\times F\times T}$, where M, F, and T denote the number of microphones, frequency bins, and time frames, respectively. Each channel's audio feature $\mathbf{X}_m$ is then encoded with a shared two-dimensional convolutional encoder, obtaining embedding $\mathbf{A}_m \in \mathbb{R}^{D \times T \times F}$ for each audio channel, where D denotes the feature dimension.

\textbf{Microphone Geometry Encoding}
To encode the array geometry, we adopt the Phase-Modulation Microphone Positional Encoding from GI-DOAEnet \cite{gidoaenet}, which first represent the microphone coordinates in spatial coordinate centered at the array center, and then project into the spectral feature dimension through a linear layer followed by layer normalization. This result in the geometry embedding as $\mathbf{g}_m \in \mathbb{R}^{D}$ shared across all time–frequency bins of microphone m.
Finally, the geometry-aware audio feature is obtained by element-wise addition between the audio and geometry encoding: $\mathbf{h}_{m,f,t} = \mathbf{A}_{m,f,t} + \mathbf{g}_m$.

\textbf{Cross-Microphone Attention Module}
The geometry-aware audio feature encodes audio information received per microphone. While the directional information of the speech is contained in the cross-channel audio info. 


To model the frequency dependent spatial information from across audio channels, while maintain model applicability of the model to varying microphone number across samples. we leverage the self-attention as the backbone model for the Cross-Microphone Interaction Module. We first introduce a learnable frequency embedding \(\mathbf{u}_f\in\mathbb{R}^{D}\) for each frequency bin \(f\), which is added to the geometry-aware microphone features:$\widetilde{\mathbf{h}}_{m,f,t} = \mathbf{h}_{m,f,t}+\mathbf{u}_f.$
The resulting features are treated as microphone tokens and processed using multi-head self-attention along the microphone dimension, allowing each microphone to interact with all others at the same frequency bin and temporal window, and preserve the T–F resolution for the downstream separation network. Since the attention parameters are shared across microphones and do not depend on their ordering or number, the encoder is permutation equivariant and supports varying microphone counts. 

\textbf{Cross-Channel Aggregation} 
Following cross-microphone interaction modeling, we employ a learnable-query attention pooling to aggregate the variable number of microphone features into a fixed-dimensional representation. Specifically, a learnable query is introduced for each frequency bin, while the cross-microphone features serve as keys and values. 
This aggregation is permutation invariant with respect to microphone ordering and supports varying numbers of microphones. We denote the aggregated audio feature as $\mathbf{a}_{aggr}$.


\subsection{Audio Extraction Backbone}

We adopt DSENet \cite{dsenet} as the speech extraction backbone. 
The azimuth angular center of the target speaker region is encoded by cyclic positional encoding, and pass through a MLP module before fusing with aggregated audio feature $\mathbf{a}_{aggr}$ through element-wise multiplication. The azimuth angular width is encoded by a BW-ConvModule, which consists of a one-hot encoder, a Linear layer, and a two dimensional convolution layer. 

The extraction backbone comprises stacked DSENet blocks, each containing a clue encoder, BW-ConvModule, Crossband layer, and Narrowband layer. The Crossband layer models local and global spectral dependencies through frequency convolutions and full-band linear transformations, respectively. The Narrowband layer captures temporal dependencies using multi-head self-attention and convolutional feedforward network along the time axis. Finally, the encoded features are projected to complex STFT coefficients, and the extracted waveform is reconstructed using the inverse STFT.

\subsection{Reconstruction Supervision Branch}
To assist the geometry-aware frontend encodes spatial and acoustic information of all the sound sources in the scene, we introduce an auxiliary multichannel audio reconstruction task. Specifically, a separated audio reconstruction model takes in the aggregated audio feature $\mathbf{a}_{aggr}$ and reconstruct the signals that would be received by a reference tetrahedral microphone array centered at the input array's geometric center with fixed orientation in the scene coordinate. 

The reconstruction decoder consists of four DSENet Crossband and Narrowband blocks, followed by linear projection to the four-channel complex STFT representation. The reconstruction branch is used only during training and is discarded during inference.

\subsection{Training Method}

Our model training include 2 stages. In the first stage, the model is trained to extract a single target speaker using spatial audio received by a 4 microphone array with varying geometry. Model is given the exact target speaker azimuth angle. All speaker lies on the same horizontal level as the microphone array, resulting in the elevation angle of 0 degrees for all speakers. Width masking module is disabled in the first stage training. In the second stage, the model is trained on microphone arrays with 4, 8 or 12 microphones. The target speaker azimuth range varies from 15, 30 and 45 degrees. The elevation of the target speakers varies between $\pm$ 20 degrees above or below the horizontal level. 0, 1 or 2 Target speakers exist in each sample, with probability of 0.3, 0.6, 0.1 respectively. 

In both training stages, we jointly optimize the extracted speech and the reconstructed tetrahedral array signals. For speech extraction, we minimize a combination of multi-resolution log-STFT magnitude loss and negative SI-SDR loss against the reverberant target speech at the reference microphone. For spatial audio reconstruction, we apply the same loss together with an inter-channel phase difference (IPD) loss to preserve spatial information. The overall objective is defined as

\[
\mathcal{L} = \mathcal{L}_{\mathrm{ext}} + \lambda_{\mathrm{rec}}\mathcal{L}_{\mathrm{rec}} + \lambda_{\mathrm{ipd}}\mathcal{L}_{\mathrm{IPD}},
\]

where \(\mathcal{L}_{\mathrm{ext}}\) and \(\mathcal{L}_{\mathrm{rec}}\) combine spectral magnitude and SI-SDR losses for the extracted and reconstructed signals, respectively. The IPD loss penalizes phase differences between microphone pairs using magnitude-weighted cosine distance. We set \(\lambda_{\mathrm{rec}}=0.25\) and \(\lambda_{\mathrm{ipd}}=20.0\). When no speaker exists in the target extraction zone, we optimize the model to output a 20 Hz sine reference tone at RMS following the design in DSENet.






\section{Experiment}
\label{sec:experiments}

\textbf{Datasets and Metrics}
We use gpuRIR \cite{gpurir} simulator to render the multi-channel spatial audio. Room size in our generation varies between 3.0 × 3.0 × 2.5 m and 10.0 × 8.0 × 4.0 m. Following GI-DOAENet \cite{gidoaenet}, we construct random microphone arrays by ensuring any 2 microphones are spaced by at least 0.01 meters and at most 0.13 meters. The receiver array is placed in the center of the room. The number of speakers vary between 2 to 6 in our data generation. Sound sources are randomly placed and are at least 0.3 meters away from the array, and 0.1 meters away from the walls. 

Samples from the LibriSpeech \cite{librispeech} dataset are used as the source signals. Additive white Gaussian noise in range 20 and 30 dB SNR with respect to the combined speech signal is used as background noise. Additional data simulation detail is in the Appendix. All the models are trained to extract the reverberant target speech received at the first microphone position input to the model. 

We calculate the extracted audio's SI-SDR, SI-SDRi, PESQ and STOI against the overlapped speech of these target speakers when one or more speakers exist in the target range. When no target speaker exists, we calculate the decay in audio intensity compared to the input audio in dB. 

\textbf{Baselines}
We compare with DSENet \cite{dsenet}, DPARNet-RSE \cite{dparn-rse}, FoVNet \cite{fovnet}, and LCMV methods on the regional target speaker extraction samples constructed as above. Since DSENet is designed to operate on a fixed microphone geometry between training and inference, we independently train two model variants. DSENet$_{Tetra}$ for the tetrahedral four-microphone array, and DSENet$_{NAO}$ for the 12-microphone array configured according to the microphone placement on the NAO robot \cite{locata}. DPARNet-RSE's architecture explicitly conditions on the microphone position information, making it support extraction across varying microphone geometries with a fixed number of microphones. Thus, we separately train DPARNet-RSE$_4$ and DPARNet-RSE$_{12}$ for the 4- and 12-microphone configurations, respectively.

To perform regional audio extraction using LCMV, we sample the target region at one degree intervals and compute the steering vector for each sampled direction at every frequency bin. These steering vectors are then used to construct the constraints of an LCMV beamformer, which is applied to the multichannel audio to extract speech from the target region.
Similarly, FoVNet beamforms over 20 predefined angular bins spanning both the target and interfering regions, converting the signal received by arrays of varying microphone number and geometry into fixed audio shape, before extract using a deep learning based backbone. We train all the baseline models by strictly following the pipelines and configurations as detailed in their paper. 

\subsection{Main Results}

\begin{table*}[]
    \centering
    \begin{tabular}{c c c cccc cccc}
    \toprule
    \multirow{2}{*}{Model} & Microphone & 0 Target & \multicolumn{4}{c}{1 Target} & \multicolumn{4}{c}{2 Target} \\
    \cmidrule(lr){3-3} \cmidrule(lr){4-7} \cmidrule(lr){8-11}
    & Geometry & Decay & SI-SDR &  SI-SDRi &  PESQ &  STOI & SI-SDR &  SI-SDRi &  PESQ &  STOI \\
    \midrule
    \multicolumn{11}{l}{\textbf{Geometry Specific Models}} \\
    \midrule
    \multirow{2}{*}{DSENet$_{Tetra}$}
    & Dynamic-4   & -23.64 & -32.61 & -26.83 & 1.09 & 0.31 & -32.56 & -37.10 & 1.08 & 0.42 \\
    & Tetrahedral & -31.64 & 5.03 & 11.36 & 1.77 & 0.75 & 10.09 & 5.63 & 2.40 & 0.84 \\
    \midrule
    \multirow{2}{*}{DSENet$_{NAO}$}
    & Dynamic-12  & -23.45 & -33.10 & -26.86 & 1.11 & 0.30 & -30.56 & -35.03 & 1.12 & 0.40 \\
    & NAO Robot   & -27.55 & 6.04 & 12.31 & 1.85 & 0.78 & 10.58 & 6.08 & 2.47 & 0.86 \\
    \midrule
    \multicolumn{11}{l}{\textbf{Geometry Aware Models (Fixed Microphone Number)}} \\
    \midrule
    \multirow{2}{*}{DPARNet-RSE$_{4}$}
    & Dynamic-4   & -23.45 & -1.10 & 4.68 & 1.29 & 0.62 & 6.63 & 2.09 & 1.80 & 0.77 \\
    & Tetrahedral & -26.54 & -2.81 & 3.52 & 1.27 & 0.61 & 6.18 & 1.72 & 1.78 & 0.77 \\    
    \midrule
    \multirow{2}{*}{DPARNet-RSE$_{12}$}
    & Dynamic-12  & -24.65 & 0.00 & 6.24 & 1.34 & 0.65 & 6.39 & 1.92 & 1.79 & 0.78 \\
    & NAO Robot   & -27.92 & -2.43 & 3.84 & 1.32 & 0.63 & 6.86 & 2.36 & 1.79 & 0.78 \\
    \midrule
    \multicolumn{11}{l}{\textbf{Geometry Aware Models (Varying Microphone Number)}} \\
    \midrule
    \multirow{4}{*}{LCMV}
    & Dynamic-4   & -3.29 & -5.87 & -0.09 & 1.12 & 0.43 & 0.45 & -4.09 & 1.35 & 0.64 \\
    & Tetrahedral & -3.56 & -6.08 & 0.25 & 1.11 & 0.43 & 0.22 & -4.23 & 1.33 & 0.63 \\
    & Dynamic-12  & -6.42 & -5.87 & 0.37 & 1.09 & 0.46 & -0.72 & -5.20 & 1.30 & 0.63 \\
    & NAO Robot   & -6.84 & -6.30 & -0.03 & 1.13 & 0.45 & -1.38 & -5.87 & 1.26 & 0.61 \\
    \midrule
    \multirow{4}{*}{FoVNet}
    & Dynamic-4   & -17.61 & -1.26 & 4.52 & 1.19 & 0.54 & 6.31 & 1.77 & 1.85 & 0.73 \\
    & Tetrahedral & -18.02 & -1.50 & 4.83 & 1.17 & 0.55 & 6.51 & 2.05 & 1.83 & 0.73 \\
    & Dynamic-12  & -19.27 & -0.94 & 5.30 & 1.20 & 0.57 & 6.63 & 2.15 & 1.84 & 0.75 \\
    & NAO Robot   & -19.29 & -1.06 & 5.21 & 1.19 & 0.56 & 6.49 & 2.00 & 1.85 & 0.73 \\
    \midrule
    \multirow{4}{*}{Ours}
    & Dynamic-4   & -26.40 & 4.01 & 9.79 & 1.65 & 0.72 & 8.80 & 4.26 & 2.22 & 0.81 \\
    & Tetrahedral & -28.24 & 4.46 & 10.79 & 1.70 & 0.73 & 9.27 & 4.82 & 2.27 & 0.82 \\
    & Dynamic-12  & -27.63 & 5.27 & 11.51 & 1.83 & 0.77 & 9.77 & 5.30 & 2.40 & 0.84 \\
    & NAO Robot   & -28.24 & 5.69 & 11.97 & 1.81 & 0.76 & 9.74 & 5.25 & 2.34 & 0.84 \\
    \midrule
    & Dynamic-4   & -24.76 & 2.98 & 8.76 & 1.47 & 0.67 & 8.16 & 3.62 & 2.09 & 0.79 \\
    Ours
    & Tetrahedral & -25.75 & 2.93 & 9.26 & 1.47 & 0.67 & 8.41 & 3.96 & 2.15 & 0.79 \\
    (FOA Recons)
    & Dynamic-12  & -26.43 & 3.11 & 9.35 & 1.55 & 0.69 & 8.90 & 4.43 & 2.20 & 0.81 \\
    & NAO Robot   & -25.89 & 3.09 & 9.36 & 1.52 & 0.68 & 8.68 & 4.18 & 2.17 & 0.79 \\
    \midrule
    & Dynamic-4   & -24.33 & 2.50 & 8.28 & 1.48 & 0.67 & 7.78 & 3.24 & 2.10 & 0.78 \\
    Ours
    & Tetrahedral & -25.06 & 2.85 & 9.18 & 1.50 & 0.68 & 8.59 & 4.14 & 2.19 & 0.81 \\
    (w/o recons)
    & Dynamic-12  & -25.99 & 3.52 & 9.76 & 1.62 & 0.72 & 8.46 & 3.98 & 2.18 & 0.81 \\
    & NAO Robot   & -26.45 & 3.96 & 10.23 & 1.61 & 0.71 & 8.90 & 4.40 & 2.22 & 0.82 \\
    \midrule
    & Dynamic-4   & -20.50 & 1.34 & 7.12 & 1.46 & 0.65 & 7.00 & 2.46 & 2.04 & 0.78 \\
    Ours
    & Tetrahedral & -21.60 & 1.29 & 7.62 & 1.45 & 0.66 & 6.82 & 2.37 & 2.05 & 0.78 \\
    (Beamform)
    & Dynamic-12  & -20.71 & 2.45 & 8.69 & 1.54 & 0.70 & 6.77 & 2.29 & 2.04 & 0.79 \\
    & NAO Robot   & -21.81 & 2.26 & 8.54 & 1.52 & 0.69 & 6.17 & 1.68 & 1.98 & 0.77 \\
    \midrule

    \end{tabular}
    \caption{Model Performance in Decay$\downarrow$, SI-SDR$\uparrow$, SI-SDRi$\uparrow$, PESQ$\uparrow$, STOI$\uparrow$ on Simulated Data across different Microphone Array Configuration and Target Speaker Number.}
    \label{tab:main-result}

\end{table*}

As shown in Table \ref{tab:main-result}, our model achieves the best performance among all geometry-aware models. Notably, increasing the number of microphones from 4 to 12 yields an approximately 1.7 dB improvement in SI-SDRi without changing the model parameters or requiring retraining. This demonstrates the ability of our model to effectively leverage additional microphone inputs and generalize across different array configurations.

Compared with geometry-specific models optimized for the tetrahedral and NAO Robot 12-microphone arrays, our model achieves slightly lower SI-SDRi scores, with performance gaps of 0.57 dB and 0.34 dB, respectively. However, these geometry-specific models exhibit poor generalization to unseen array configurations, yielding negative SI-SDRi scores on both Dynamic-4 and Dynamic-12 arrays. In contrast, our model maintains robust performance across varying microphone geometries, highlighting its ability to generalize beyond fixed array configurations while achieving performance competitive with geometry-specific models.

\subsection{Additional Evaluations}

\textbf{Ablation on Geometry-Aware Encoding Frontend Design} In this section, we compare with alternative audio encoding strategies that support microphone geometry awareness and accommodate varying numbers of microphones.

Following the design of FoVNet, we implement a beamforming-based frontend that divides the target speaker region into $N=20$ fixed angular bins and performs beamforming towards the center of each bin. The resulting beamformed signals are treated as 20 audio channels and fed into the subsequent DSENet extraction backbone. As an alternative, we employ a least-squares mode-matching algorithm to transform the multichannel microphone signals into a four-channel first-order Ambisonics (FOA) representation, which is then used as input to DSENet. Both variants are trained using the same optimization strategy as our proposed model.

As shown in Table \ref{tab:main-result}, both approaches demonstrate the ability to generalize across microphone geometries and array sizes, achieving positive SI-SDRi scores on both Dynamic-4 and Dynamic-12 configurations. However, both consistently underperform our proposed method across all evaluation scenarios and metrics. These results highlight the effectiveness of our geometry-aware encoding frontend in encoding spatial information from the geometry varying array.

\textbf{Ablation on Reconstruction Supervision Signal} We further investigate the contribution of the auxiliary reconstruction objective by removing the tetrahedral microphone array signal reconstruction loss during training.

As shown in Table \ref{tab:main-result}, removing this objective leads to a consistent reduction of approximately 1.6 dB in SI-SDRi across all evaluated microphone configurations. This demonstrates the effectiveness of the auxiliary reconstruction supervision in encourages the model to learn more informative spatial representations.

\textbf{Performance on Real Audio Recordings}
\begin{table}[]
    \centering
    \begin{tabular}{c cc cc}
    \toprule
    \multirow{2}{*}{Model} & \multicolumn{2}{c}{Tetrahedral} & \multicolumn{2}{c}{12 Mic NAO Robot} \\
    \cmidrule(lr){2-3} \cmidrule(lr){4-5}
    & simulated & real & simulated & real \\
    \midrule
    DSENet & 11.36 & 8.40 & 12.31 & 9.57 \\
    DPARNet-RSE & 3.52 & 5.19 & 3.84 & 6.11 \\
    LCMV & 0.25 & 1.12 & -0.03 & -2.58 \\
    FoVNet & 4.83 & 3.85 & 5.21 & 4.96 \\
    Ours & 10.79 & 9.15 & 11.97 & 9.08 \\

    \midrule
    \end{tabular}
    \caption{Model Performance comparison between Simulated and Real Data under single target speaker scenario.}
    \label{tab:real-data}
\end{table}
In this section, We further evaluate the generalization of different methods to real-world recordings. We comparing the performance on real and simulated audio captured using the same microphone array geometries. All models are trained exclusively on simulated data, allowing us to assess their robustness to the simulation-to-reality (sim-to-real) gap.

As shown in Table \ref{tab:real-data}, our model achieves an SI-SDRi of 9.15 dB on real tetrahedral microphone recordings, corresponding to a 1.64 dB decrease compared with simulated recordings. In contrast, the geometry-specific model exhibits a larger sim-to-real performance gap of 2.96 dB, achieving only 8.40 dB SI-SDRi on real recordings. This indicates that our method is more robust to the sim-to-real gap for the tetrahedral array, despite its slightly lower performance on simulated data.

For the NAO Robot 12-microphone array, the geometry-specific model achieves 9.57 dB SI-SDRi on real recordings, with a sim-to-real performance gap of 2.74 dB. In comparison, our method achieves 9.08 dB SI-SDRi, with a comparable sim-to-real gap of 2.89 dB. These results indicate that the relative robustness to the sim-to-real gap varies across microphone configurations. While our method demonstrates better generalization to real recordings on the tetrahedral array, it achieves comparable sim-to-real robustness on the NAO Robot array, despite a modest reduction in absolute extraction performance.

\textbf{Evaluation on Extract-Remove Boundary}
Finally, we investigate the model's ability to accurately extract or suppress speakers located near the boundaries of the target region. We introduce a gain-based metric to quantify boundary extraction accuracy and empirically examine how microphone array geometry influences this performance.

We focus our evaluation on two-speaker mixtures, where one interfering speaker is always located outside the target region, while a probe speaker is positioned near the target region boundary at angular offsets of $\delta = \pm 2^\circ$. Negative offsets indicate that the probe speaker lies inside the target region, whereas positive offsets indicate that it lies outside.


To quantify the model's ability to extract or suppress the probe speaker, we define a gain-based metric. Given the model output $\hat{y}$ and the reverberant speech signal $s$ of the probe speaker, the gain is calculated as $\alpha(\hat{y}) = \frac{\hat{y}^{T}s}{|s|^2}$. Ideally, perfect extraction of the probe speaker yields $\alpha(s)=1$, while complete suppression yields $\alpha(\mathbf{0})=0$. We therefore adopt a decision threshold of $\tau=0.5$, which is at equidistant from the both ideal cases.

We evaluate boundary extraction performance using balanced accuracy, defined as $Acc_{\mathrm{Balanced}}(d) =\frac{Acc_{\mathrm{extract}}(d)+Acc_{\mathrm{remove}}(d)}{2}$, where $Acc_{\mathrm{extract}}(d) =\mathbb{E}[\mathbf{1}(\alpha(\hat{y})>\tau)\mid\delta < 0]$, 
$Acc_{\mathrm{remove}}(d) =\mathbb{E}[\mathbf{1}(\alpha(\hat{y})<\tau)\mid\delta > 0]$
. This metric equally weights the model's ability to retain speakers inside the target region and suppress those outside.

Using the tetrahedral microphone array, we observe that the proposed model tends to retain speech from speakers located just outside the target region, resulting in limited boundary discrimination accuracy. Specifically, the model achieves balanced accuracies of 0.500, 0.500, and 0.589 for target region widths of $15^\circ$, $30^\circ$, and $45^\circ$, respectively. In comparison, the geometry-specific DSENet achieves substantially higher balanced accuracies of 0.844, 0.875, and 0.821 under the same conditions. These results reveal a limitation of the proposed geometry-aware model in precisely distinguishing speakers near the target region boundary when using the tetrahedral array.

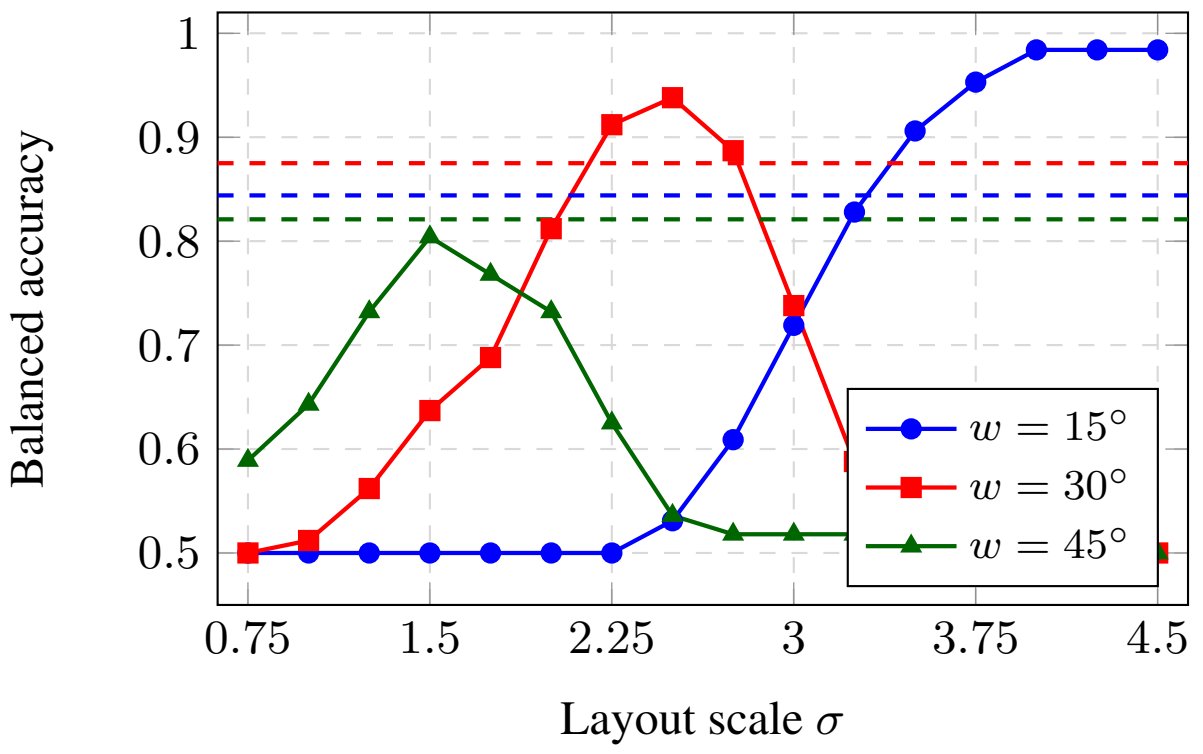
\begin{figure}[ht]
    \centering
    
    \begin{tikzpicture}
    \begin{axis}[
        width=8cm, height=5.5cm,
        xlabel={Layout scale $\sigma$},
        ylabel={Balanced accuracy},
        xmin=0.625, xmax=4.625, ymin=0.45, ymax=1.02,
        xtick={0.75,1.5,2.25,3.0,3.75,4.5},
        ytick={0.5,0.6,0.7,0.8,0.9,1.0},
        grid=major, grid style={dashed,gray!30},
        legend pos=south east, legend cell align=left,
        legend style={font=\footnotesize}, mark size=1.6pt,
    ]
    \addplot[color=blue, mark=*, thick] coordinates {
        (0.75,0.500) (1.00,0.500) (1.25,0.500) (1.50,0.500) (1.75,0.500) (2.00,0.500)
        (2.25,0.500) (2.50,0.531) (2.75,0.609) (3.00,0.719) (3.25,0.828) (3.50,0.906)
        (3.75,0.953) (4.00,0.984) (4.25,0.984) (4.50,0.984)
    };
    \addlegendentry{$w=15^\circ$}
    \addplot[color=red, mark=square*, thick] coordinates {
        (0.75,0.500) (1.00,0.512) (1.25,0.562) (1.50,0.637) (1.75,0.688) (2.00,0.812)
        (2.25,0.912) (2.50,0.938) (2.75,0.887) (3.00,0.738) (3.25,0.588) (3.50,0.550)
        (3.75,0.512) (4.00,0.512) (4.25,0.500) (4.50,0.500)
    };
    \addlegendentry{$w=30^\circ$}
    \addplot[color=black!60!green, mark=triangle*, thick] coordinates {
        (0.75,0.589) (1.00,0.643) (1.25,0.732) (1.50,0.804) (1.75,0.768) (2.00,0.732)
        (2.25,0.625) (2.50,0.536) (2.75,0.518) (3.00,0.518) (3.25,0.518) (3.50,0.500)
        (3.75,0.500) (4.00,0.500) (4.25,0.500) (4.50,0.500)
    };
    \addlegendentry{$w=45^\circ$}

    \addplot[color=blue, dashed, thick, domain=0.625:4.625, samples=2, forget plot] {0.844};
    \addplot[color=red, dashed, thick, domain=0.625:4.625, samples=2, forget plot] {0.875};
    \addplot[color=black!60!green, dashed, thick, domain=0.625:4.625, samples=2, forget plot] {0.821};
    \end{axis}
    \end{tikzpicture}
    \caption{Balanced Accuracy under different microphone aparture scaling and target speaker angular width. Dashed lines indicate the Balanced Accuracy of the geometry specific DSENet model at each target speaker angular width.}
    \label{fig:boundary}
\end{figure}

We further investigate whether modifying the microphone array geometry can improve boundary extraction accuracy. Specifically, we scale the aperture of the tetrahedral array along the direction perpendicular to the center of the target region by a factor of $\sigma$.

As shown in Figure \ref{fig:boundary}, increasing the array aperture generally improves boundary discrimination for narrower target regions. For a target region width of $15^\circ$, the highest balanced accuracy is achieved at $\sigma=4.0$, while $\sigma=2.5$ yields the best performance for a width of $30^\circ$. In both cases, the optimized geometries outperform not only the original tetrahedral configuration of our geometry-aware model but also the geometry-specific DSENet baseline. For the wider $45^\circ$ target region, the improvement is less pronounced, with the highest balanced accuracy achieved at $\sigma=1.5$. Visualizations of the scaled microphone configurations are provided in the Appendix.

\begin{figure}[!t]
    \centering
    \subfigure{
        \includegraphics[width=0.305\linewidth]{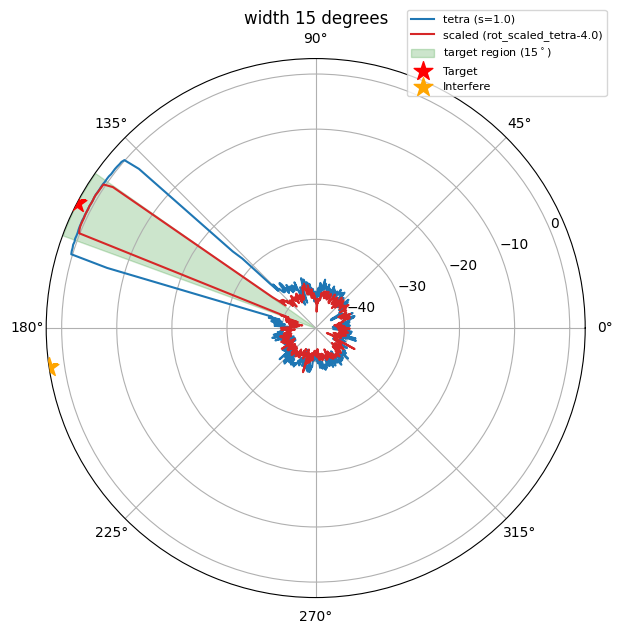}
    }
    \subfigure{
        \includegraphics[width=0.305\linewidth]{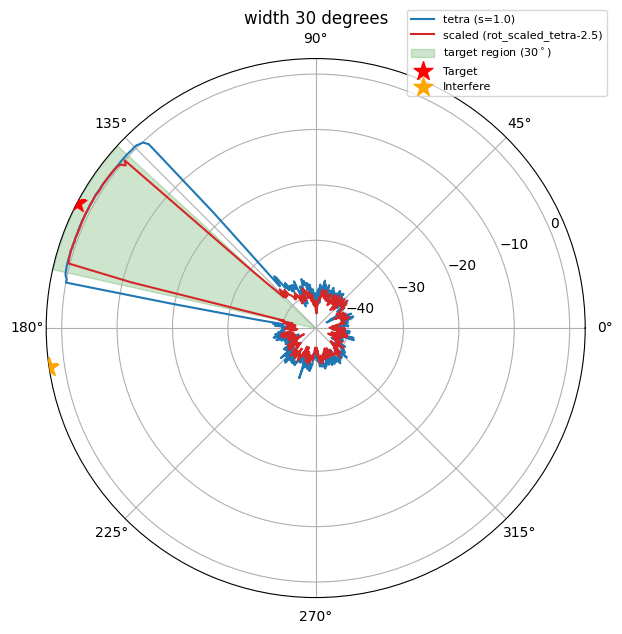}
    }
    \subfigure{
        \includegraphics[width=0.305\linewidth]{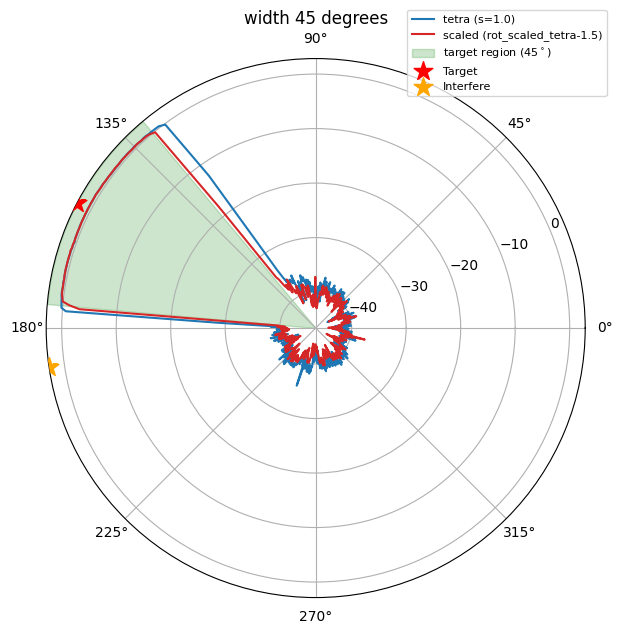}
    }
    \caption{Visualization of model extraction performance before and after scaling the microphone array aperture size.}
    \label{fig:boundary_visualize}
\end{figure}

In Figure \ref{fig:boundary_visualize}, we visualize the model angular decay distribution in polar coordinate. We fix the target region and evaluate the model with the probe speaker placed at 500 equally spaced azimuth angles around the microphone array, maintaining a constant source-to-array distance. At each position, we calculate the decay of the extracted signal relative to the probe speaker's reverberant speech and visualize the resulting angular decay distribution.

Compared with the original tetrahedral array, the optimized configurations with larger apertures produce angular decay distribution that more closely follow the prescribed target region boundaries, indicating improved spatial selectivity and more accurate suppression of speakers outside the target region. These results demonstrate that the proposed geometry-aware model can leverage microphone geometry optimization to improve boundary extraction accuracy without retraining, highlighting an advantage over geometry-specific models that are optimized for fixed microphone array configurations.

\section{Conclusion}

In conclusion, we propose a microphone geometry-aware model to perform regional target speaker extraction. We demonstrate our model's generalizability across different microphone array geometries, while maintaining under 1dB SI-SDRi performance gap with the model trained on geometry specific data. We further analysis the model extract and suppress accuracy at the specified target speaker location boundaries.


\bibliographystyle{IEEEtran}
\bibliography{6_reference}


\end{document}